\documentclass[11pt,a4paper]{article}
\usepackage{jstyle}

\usepackage[utf8]{inputenc}
\usepackage{amsthm,amsmath,latexsym,amssymb,amsfonts,amssymb,amscd}
\usepackage{hyperref}
\usepackage{cancel}
\usepackage{color}
\usepackage{mathabx}

\definecolor{rougef}{rgb}{0.56,0,0}
\definecolor{vertf}{rgb}{0,0.5,0}
\definecolor{bleuf}{rgb}{0,0,0.8}
\definecolor{violetf}{rgb}{0.5,0,0.5}

\author{\quad Karapet MKRTCHYAN}

\affiliation{Max Planck Institute for Gravitational Physics (Albert Einstein Institute)\\
Am M\"uhlenberg 1, 14476 Potsdam, Germany}

\emailAdd{karapet.mkrtchyan@aei.mpg.de}

\title{\centering
\LARGE{Cubic interactions of massless bosonic fields in three dimensions}}

\abstract{Parity-even cubic vertices of massless bosons of arbitrary spins in three dimensional Minkowski space are classified in the metric-like formulation. As opposed to higher dimensions, there is at most one vertex for any given triple $s_1,s_2,s_3$ in three dimensions. All the vertices with more than three derivatives are of the type $(s,0,0)$, $(s,1,1)$ and $(s,1,0)$  involving scalar and/or Maxwell fields. All other vertices contain two (three) derivatives, when the sum of the spins is even (odd). Minimal coupling to gravity, $(s,s,2)$, has two derivatives and is universal for all spins (equivalence principle holds). Minimal coupling to Maxwell field, $(s,s,1)$, distinguishes spins $s\leq 1$ and $s\geq 2$ as it involves one derivative in the former case and three derivatives in the latter case. Some consequences of this classification are discussed.}

\begin{document}

\maketitle

\section{Introduction}

When trying to understand several complicated phenomena in physics, it is useful to start from the spherical cows in the vacuum. For higher-spin (HS) theories  \cite{Fierz,Fronsdal:1978rb,Vasiliev:1990en,Prokushkin:1998bq,Vasiliev:2003ev,bciv,Didenko:2014dwa,Didenko:2017qik} --- field theories containing massless fields with spins $s>2$ --- these cows are three dimensional models. They have attracted considerable interest during recent years (see e.g. \cite{Campoleoni:2012hp,Fredenhagen:2014oua,Campoleoni:2014tfa,Campoleoni:2017xyl})
due to their simplicity, nevertheless encoding non-trivial physics \cite{Banados:1992wn,Campoleoni:2010zq}.

Three dimensional higher spin theories have been useful laboratories for simple models of holography \cite{Gaberdiel:2010pz,Ammon:2011ua,Gaberdiel:2012uj,Ammon:2011nk} and explorations of physics beyond the spin two horizon \cite{Ammon:2011nk,Gwak:2015vfb,Gwak:2015jdo,Joung:2017hsi}.
The possibility to formulate them as Chern-Simons theories \cite{Achucarro:1987vz,Blencowe:1988gj}, \cite{Campoleoni:2010zq,Gwak:2015vfb,Gwak:2015jdo} with non-compact gauge groups makes general relativity and (colored) HS theories in three dimensions a simple playground for tackling problems relevant to quantum gravity. This simplicity is lost, however, as soon as one adds matter into the picture \cite{Prokushkin:1998bq,Kessel:2015kna}. On the other hand, the HS theory with matter content \cite{Prokushkin:1998bq} is more similar to the situation in higher dimensions \cite{Vasiliev:1990en}.
Not surprisingly, holographic models of higher spins in three dimensions also require matter content \cite{Gaberdiel:2010pz,Gaberdiel:2012uj} and therefore their understanding requires careful study of the matter coupling in HS systems in three dimensions \cite{Prokushkin:1998bq,Ammon:2011ua}.  In a sense, once matter is added, the three-dimensional theory becomes almost as complicated as its four dimensional counterpart \cite{Vasiliev:1990en}. Nevertheless, we will see in this work, that the cubic interactions of HS fields in metric-like formulation are totally different from those in higher dimensions and have certain inner simplicity, possibly inherited from Chern-Simons formulation.

The question of the Lagrangian formulation of HS theories is a long standing puzzle, addressed in particular through attempts for perturbative constructions of the action for HS theories, in the spirit of the so-called Fronsdal program \cite{Fronsdal:1978rb}, that resulted so far in full classifications of cubic interactions in dimensions higher than three (see e.g.
\cite{Manvelyan:2010jr,Manvelyan:2010je,Joung:2011ww,Conde:2016izb,Metsaev:2012uy,Joung:2012fv,Francia:2016weg}).

In three dimensions, unlike higher dimensions, HS interacting theories can be considered not only in constant non-zero curvature backgrounds, but also in Minkowski space, since the main obstacles in higher dimensions --- Weinberg's S-matrix argument \cite{Weinberg:1964ew} and Aragone-Deser problem \cite{Aragone:1979hx} are not relevant in $d=3$. The first one is not relevant due to the fact, that there are no HS propagating particles in three dimensions, while the second one does not apply as shown by Aragone and Deser themselves \cite{Aragone:1983sz} (see also \cite{Campoleoni:2011tn,Zinoviev:2014sza}) due to the admissibility of minimal gravitational coupling in Minkowski space of three dimensions. Together with the fact that the Fronsdal program is technically simpler in Minkowski space compared to (A)dS, this makes the three dimensional flat space a preferred playground for the problem of Lagrangian formulation for non-linear HS theories with propagating matter content.

Still, the systematics of three-linear interactions in $d=3$ is not yet completely known. Indeed, there are only a handful of works on higher spin interactions in three dimensions in the metric-like formulation (see e.g. \cite{Campoleoni:2012hp,Fredenhagen:2014oua,Campoleoni:2014tfa}).

In this work, we start an investigation in this direction, proposing a classification of cubic vertices in three dimensions. Let us stress, that in $d=3$ the on-shell S-matrix methods \cite{Benincasa:2007xk,Conde:2016izb} or light-cone technologies \cite{Bengtsson:1986kh,Metsaev:2005ar,Metsaev:1991mt,Metsaev:1991nb,Ponomarev:2016lrm,Ponomarev:2016cwi,Ponomarev:2017nrr} do not apply for massless fields with $s\geq 2$, while the higher dimensional covariant classification \cite{Manvelyan:2010jr} has little overlap with the three dimensional one, as it will be clear from our analysis.

The main technical difference between the three-dimensional and the higher-dimensional classifications of cubic vertices for HS fields is that in $d=3$ there exist dimension dependent identities (Schouten identities), that should be taken into account in the cohomological problem (see e.g. \cite{Barnich:2000zw}) of finding cubic vertices.

In the case when there are no Schouten identities, the problem of cubic  deformations can be formulated as finding a functional, linear in all three fields, that:
\begin{enumerate}
\item
is not in the same equivalence class of total derivatives and terms proportional to the free equations of motion (the latter terms are induced by field redefinitions in the quadratic action) with zero:
\begin{align}
\mathcal{L}_3\approx \mathcal{L}_3+\partial_{\mu}J^{\mu}+ \sum_i \Delta_i \cF_i\,,\napprox 0
\end{align}
where $\cF_i$ denote the free equations of motion of the fields involved.

\item
its variation with respect to the gauge symmetry of each field vanishes {\it in the same equivalence class}, i.e. up to total derivatives and terms proportional to the free equations of motion (these  terms induce deformations of gauge transformations):
\begin{align}
\delta_i^{(0)}\mathcal{L}_3\approx 0\,,
\end{align}
\end{enumerate}

Differently, in the presence of Schouten identities, the problem gets reformulated as the question of finding cubic functionals such that:
\begin{enumerate}
\item
are not equivalent to zero in equivalence class of total derivatives, free equations of motion {\it and Schouten identities},
\begin{align}
\mathcal{L}_3\bar{\approx} \mathcal{L}_3+\partial_{\mu}J^{\mu}+ \sum_i \Delta_i \cF_i+\sum_k \bar\Delta_k \mathcal{D}_k\napprox 0\,,
\end{align}
where $\mathcal{D}_k$ are all the Schouten identities in given dimension.
\item
their variation with respect to the gauge symmetry of each field vanishes {\it in the same equivalence class}, i.e. up to total derivatives, terms proportional to the free equations of motion {\it and Schouten identities.}:
\begin{align}
\delta_i^{(0)}\mathcal{L}_3\bar\approx 0\,,
\end{align}
\end{enumerate}

In all cases, one has a well defined cohomological problem, but depending on the content of the Schouten identities the solution may be different.
Solutions that do not take the Schouten identities into account may be identified as trivial solutions once Schouten identities are taken into account, therefore cannot be trusted.

The relevance of the properly considered Schouten identities is related to the fact that we work in Lorentz covariant basis that hides the triviality of certain expressions.
In six and higher dimensions there are no Schouten identities relevant to cubic vertices of symmetric fields, while in five dimensions, the only Schouten identity is a total derivative. 

Taking the Schouten identities into account has shown to result in less vertices in four dimensions compared to those in higher dimensions \cite{Conde:2016izb}.
An illustrative example in four dimensions is the cubic term in the linearised expansion of Gauss-Bonnet action. It is a gauge invariant vertex in any dimensions, but turns out to be a total derivative in four dimensions, up to a Schouten identity.
As we will show in this work, the consequences in $d=3$ are even more drastic.

As shown in \cite{Manvelyan:2010jr}, the cubic vertices for Fronsdal fields can be constructed order by order in traces and divergences of the fields involved, starting from the piece that does not involve any trace or divergence, which is usually referred to as traceless-transverse (TT) vertex. As demonstrated in \cite{Francia:2016weg}, these leading pieces of interactions are the same for Fronsdal and Maxwell-like \cite{SV,Campoleoni:2012th,FLS,Francia:2016weg} formulations of free HS theory. In a sense, the TT classification is universal and provides the first step towards the off-shel cubic action for both Fronsdal and Maxwell-like HS fields. The TT cubic vertex, in a sense, is solving the problem of quadratic deformation of the wave equations for Fierz, i.e. TT fields \cite{Fierz}. In this paper we will work at the level of TT fields, without making any distinction between Fronsdal fields and Maxwell-like fields, bearing in mind that the TT vertices found here can be completed to off-shell vertices in both approaches. Therefore we do not solve the problem of finding off-shell vertices, but rather that of classifying them. In a sense, the results of this paper can be regarded as the three-dimensional analogue of the light-cone classification in higher dimensions \cite{Metsaev:2005ar}.

The paper is organised as follows. In Section \ref{CubicRev} we start by a lightning  review of the covariant classification of cubic vertices for massless symmetric fields in $d\geq 4$. In Section \ref{DDI3d} we start the investigation of three-dimensional vertices by first deriving special identities related to cubic vertices that hold only in $d\leq 3$.
In Section \ref{LowSpin} we derive cubic vertices involving lower spin fields. These include all interactions involving scalar and Maxwell fields, as well as gravitational couplings for any spin. In Section \ref{General} we derive the full classification of vertices involving any spins $s_1\geq s_2\geq s_3$. We conclude in Section \ref{Discussion} with some discussion on the implications and on the potential use of our results.

\section{Review of Cubic Vertices in Higher Dimensions}\label{CubicRev}

We will review here the classification \cite{Manvelyan:2010jr} of cubic vertices in dimensions higher than three, which allows us to introduce notations and conventions and set the stage for interactions in three dimensions. We will follow simple notations of \cite{Joung:2012fv,Conde:2016izb}.

Spin $s$ massless field in three dimensions is parametrised by a symmetric $s$-th rank tensor $\phi_{\mu_1\dots\mu_s}$. We will contract all the indices of these fields with auxiliary vector variables, $a^\mu$,
to make the symmetry of indices manifest, as well as make the expressions for the Lagrangian scalar terms more compact, hiding the complexity of index contractions.
\begin{align}
\phi^{(s)}(x, a)=\frac1{s!}\phi_{\mu_1\dots\mu_s} a^{\mu_1}\cdots a^{\mu_s}\,.\label{phi}
\end{align}
Lorentz covariant formulation of massless fields with spin require gauge symmetries, realised as gradient transformations,
\begin{align}
\delta_{\epsilon}^{(0)}\phi^{(s)}(x, a)=(a\cdot \partial)\epsilon^{(s-1)}(x, a)\,,\\
\epsilon^{(s-1)}(x, a)=\frac1{(s-1)!}\epsilon_{\mu_1\dots\mu_{s-1}}\,.
\end{align}
The most conventional formulation of the free theory of massless symmetric fields is the Fronsdal formulation \cite{Fronsdal:1978rb}, where the fields $\phi$ are double traceless, while the parameters $\epsilon$ are traceless.

We will follow the logic of Noether procedure, assuming existence of action principle of the full interacting theory, and the Lagrangian can be expanded in a small parameter $g$:
\begin{align}
\mathcal{L}=\mathcal{L}^{(2)}+g \mathcal{L}^{(3)}+O(g^2)
\end{align}
The cubic action is a sum of different vertices
\begin{align}
\mathcal{L}^{(3)}=\sum_{s_i,n}g_{s_1,s_2,s_3}^n\mathcal{L}^n_{s_1,s_2,s_3}\,,
\end{align}
for collections of three fields with arbitrary spins $s_1\geq s_2\geq s_3$, and $n$ is a parameter that counts independent basis of vertices (number of free parameters, that are not fixed by the gauge invariance of the interaction between given three fields). At the cubic order, the existence of a certain vertex does not depend on the full spectrum of the theory, and the coupling constants in front of each vertex are completely arbitrary. Of course, due to the requirement of gauge invariance, the number of vertices are much less then the number of Poincar\'e invariant vertex monomials one can write. For example, in $d\geq 5$, this requirement leaves one overall coefficient for each number of derivatives in the allowed range $s_1+s_2-s_3, s_1+s_2-s_3+2,\dots,s_1+s_2+s_3$. The lowest and highest possible number of derivatives were established first in light cone classification \cite{Metsaev:2005ar} and are called Metsaev bounds. As we will see in the following, in three dimensions for each triple of spins $(s_1,s_2,s_3)$ there is at most one vertex.

Cubic vertex monomials can be written in a compact form, once the building blocks of them are classified. These building blocks are operators of scalar contractions between the fields $\phi_i(x_i,a_i)$, and their derivatives, given by all possible contractions of the operators
 $\partial_{a_i}^{\mu}=\frac{\partial}{\partial a_\mu}$ and $\partial^i_{\mu}=\frac{\partial}{\partial x_i}$. They can be classified into following groups:
\begin{align}
B_{ij}=\partial_i\cdot\partial_j\,,\quad y_i=\partial_{a_i}\cdot\partial_{i+1}\,,\quad z_i=\partial_{a_{i+1}}\partial_{a_{i-1}} \,,\quad Div_i=\partial_{a_i}\cdot\partial_i\,,\quad Tr_i=\partial_{a_i}\cdot\partial_{a_i}\,.
\end{align}
At the end we have a local functional in the action, therefore have to integrate over $dx_i \delta(x-x_i)$. The splitting of the coordinates is useful for keeping track over the derivatives acting on different fields, but is just a trick and has no physical consequence related to locality.
Also, the Lagrangian monomials are Lorentz scalars, which means that at the end they do not depend on auxiliary variables used to write them, so we can evaluate all the $a_i$ to zero.
Note, that choice of variables already fixed the partial integration freedom, since we chose to not use linearly dependent operators $\partial_{a_i}\cdot \partial_{i-1}\approx -y_i-Div_i$, since $\partial_1^\mu+\partial_2^\mu+\partial_3^\mu\approx 0$.

When discussing only the TT part of the vertex, we can discard all the terms proportional to divergences and traces of the fields, that are given by the operators $Div_i$ and $Tr_i$. We can also fix the field redefinition freedom, requiring that in the TT part of the vertex there are no operators of the type $B_{ij}$\footnote{It is useful to note, that up to total derivatives, $B_{ii+1}=\frac12(\partial_{i-1}^2-\partial_{i}^2-\partial_{i+1}^2)$, therefore any $B_{ij}$ can be represented through Laplacian operators acting on the fields, which in turn can be traded with divergence and trace terms using field redefinition in the free action (or free equations of motion).}. After all, all the vertices can be written in terms of variables $y_i, z_i$.
\begin{align}
\mathcal{L}^{n}_{s_1,s_2,s_3}=\mathcal{V}^n_{s_1,s_2,s_3}(y_i,z_i)\phi_1(x_1,a_1)\phi_2(x_2,a_2)\phi_3(x_3,a_3)\,,
\end{align}
where vertex operator $\mathcal{V}^n_{s_1,s_2,s_3}$ is a polynomial in variables $y_i, z_i$.
It is easy to show, that once we fix the field redefinition freedom as described above, TT vertex, containing  two identical fields is either symmetric or antisymmetric in their exchange, depending on the spin of the third field. 

Gauge variation of the field $\phi_i$, $\delta\phi_i=a_i\cdot \partial_i\,\epsilon_i$, is given by action of a certain operator on the vertex. In order to derive this operator, we first note that after gauge variation we will need to compute commutators of gradient operators $a_i\cdot \partial_i$ with $y_j$ and $z_j$.
These commutators are given as:
\begin{align}
[y_i,a_i\cdot\partial_i]=B_{ii+1}=0\,,\quad [y_{i\pm 1},a_i\cdot\partial_i]=0\,,\\
 [z_i,a_i\cdot\partial_i]=0\,,
\quad [z_{i\pm 1},a_i\cdot\partial_i]=\pm y_{i\mp 1}\,,
\end{align}
therefore, the operator of gauge variation, acting on the vertex is:
\begin{align}
D_i \mathcal{L}^{(3)} \equiv (y_{i-1} \partial_{z_{i+1}}- y_{i+1} \partial_{z_{i-1}}) \; \mathcal{L}^{(3)} =0,\quad \forall i \in \{1,2,3\}\label{Var}
\end{align}
When the two identical fields (say, $s_1=s_2$) are exchanged in all monomials of the vertex, the monomials all pick up the same sign factor $(-1)^{s_3}$. Then it is straightforward to see that the variation of the non-zero vertex with respect to two identical fields will be identical, so vanishing of their sum implies vanishing of separate variations of each of fields. Therefore, the equation \eqref{Var} is complete.
In the case if $s_3$ is odd, the non-trivial vertex requires Chan-Paton factors, that allow for the first and second fields to have different quantum numbers.

The solution to \eqref{Var} in any dimensions is given by the following expression:
\begin{align}
\mathcal{L}^{(3)}=\mathcal{V}(y_i, G)\phi_1\phi_2\phi_3\,,\quad G=y_1z_1+y_2z_2+y_3z_3\,.
\end{align}
For given spins $s_1\geq s_2\geq s_3$, we have:
\begin{align}
\mathcal{V}^n_{s_1,s_2,s_3}=g^n_{s_1,s_2,s_3} y_1^{s_1-n}y_2^{s_2-n}y_3^{s_3-n}G^n\,,\quad n=0,1,\dots,s_3\,.
\end{align}
In four dimensions, there is a Schouten identity (see \cite{Conde:2016izb} for more details)
\begin{align}
y_1y_2y_3G=0\,,\label{GBDDI}
\end{align}
which is responsible for vanishing of all the vertices with $0<n<s_3$. The simplest example is Gauss-Bonnet vertex
\begin{align}
\mathcal{V}_{GB}=y_1y_2y_3G\,.
\end{align}
The four dimensional Schouten identity \eqref{GBDDI} is of course also a Schouten identity in three dimensions, but in three dimensions there are more Schouten identities. We will derive systematically the fundamental Schouten identities in the next section, and use in classification of cubic vertices in three dimensions.

\section{Schouten identities and Three dimensional Vertices}\label{DDI3d}

Since in the following we will concentrate on TT vertex, we define a new equivalence class, that is equivalence up to terms, proportional to total derivatives, operators $B_{ij}, Div_i, Tr_i$ and all the Schouten identities to be derived in the following. We will completely discard the terms, proportional to all of these operators in the following and use equality sign $=$ between two vertex operators that are equivalent modulo these operators. One has to bear in mind that this way we will only track TT terms. Once the TT vertices are known, one can use the procedure of \cite{Manvelyan:2010jr,Manvelyan:2010je,Francia:2016weg} for order by order construction of off-shell vertex. We will leave that for further work.

Dimensional dependent identities (Schouten identities) are constructed by 'over-antisymmetrization', i.e. by contracting arbitrary tensors with
\begin{align}
\delta^{\mu_1 \dots \mu_{d+1}}_{\nu_1 \dots \nu_{d+1}} \equiv 0 \,,
\end{align}
where $d$ is the space-time dimension.
In this manner, all Schouten identities can be systematically derived. Such identities allow for the existence of additional gauge invariant vertices, namely those which obey
\begin{align}
 D_i \mathcal{V} \equiv (y_{i+2} \partial_{z_{i+1}}- y_{i+1} \partial_{z_{i+2}}) \; \mathcal{V} = \text{ Schouten identity terms},\quad \forall i \in \{1,2,3\} \label{CVE}
\end{align}

These vertices are gauge invariant for the dimension of interest since the Schouten identities vanish identically.

In dimensions higher than four, Schouten identities are to be created by contracting at least six antisymmetric indices. This kind of contraction necessarily gives a total divergence term that is irrelevant in our discussion since we discard boundary terms. Therefore, there are no Schouten identities relevant to cubic interactions of symmetric fields in $d\geq 5$. As was mentioned before, in $d=4$, many covariant cubic vertices  are proportional to Schouten identities \cite{Conde:2016izb}. There, those vertices get killed, with no others appearing in return. As we will show in the following, the situation is different in three dimensions --- we have large class of cubic vertices killed by Schouten identities and another large class of non-trivial vertices that are gauge invariant only up to Schouten identities, therefore are intrinsically three-dimensional.

In the following, we will focus on the three-dimensional case. The complete list of parity even elementary Schouten identities is then given by
\begin{subequations}
\begin{align}
(G - y_i z_i)^2 = 0 \,, && y_i z_i G - y_{i-1} z_{i-1} y_{i+1} z_{i+1} = 0 \,, \label{eq:twoDerivDDI}\\
y_i y_{i\pm 1}(G - y_i z_i) = 0 \,, \label{eq:threeDerivDDI}\\
y_i^2 y^2_{i+1} = 0 \,, && y_i^2 y_{i+1} y_{i-1}=0 \label{eq:fourDerivDDI}\,.
\end{align} 
\end{subequations}
Note that we have grouped the identities in two-, three- and four- derivative expressions. Any other Schouten identity, relevant to our problem can be derived from these elementary ones. For example, the Schouten identity \eqref{GBDDI} can be found multiplying \eqref{eq:threeDerivDDI} by $y_{i\mp 1}$ and taking into account \eqref{eq:fourDerivDDI}.
One can derive many useful consequences. A useful example is: $y_i^4 z_i^2=0$. We will use all of these identities in the following.

\section{Lower spin examples}\label{LowSpin}

\subsection{Vertices with scalars}\label{Scalar}

To start with, we take the simplest example, where two of the fields are scalars ($s_2=s_3=0, s_1=s$). It is straightforward to see, that the only Lorentz-invariant expression we can form in this case, is:
\begin{align}
\mathcal{V}_{s,0,0} =  y_1^{s}\,, \label{s00}
\end{align}
and coincides with generic-dimensional expression (see e.g. \cite{Manvelyan:2009tf,Bekaert:2009ud}). Properties of the conserved currents, associated to the extension of this vertex to $(A)dS_3$ space, was discussed in \cite{Prokushkin:1999xq}.

Next we turn to the case where we have only one scalar field in the cubic vertex ($s_3=0, s_1\geq s_2\geq 1$).
It can be shown using the Schouten identities, that for $s_2\geq 2$ there is no vertex of this type, while for $s_2=1$ there is a unique vertex with $s_1+1$ derivatives:
\begin{align}
\mathcal{V}_{s_1,1,0} = y_1^{s_1}y_2\,,\label{s10}
\end{align}
with deformation of gauge transformation for scalar field, unless $s_1=1$.

We find that for the vertices involving scalar fields, the difference in three dimensions as compared to higher dimensions is the absence of vertices of interactions of the scalar with two fields of spins $s_1\geq s_2\geq 2$. The only vertices are $(s,0,0)$ \eqref{s00} and $(s,1,0)$ \eqref{s10}, which are given by the same expression as in higher dimensions.

\subsection{Vertices with Maxwell fields}\label{Maxwell}

Now we turn to the next simple example --- vertices with Maxwell fields ($s_3=1, s_1\geq s_2\geq 1$).
We find two non-trivial solutions in this case.

One solution works for $s_2=1$, and is given by the expression that works in any dimensions --- current coupling, studied e.g. in \cite{Mkrtchyan:2010pp,Manvelyan:2009vy,Mkrtchyan:2010zz}.
\begin{align}
\mathcal{V}_{s,1,1} = y_1^{s-1}G\,,\label{s-1-1}
\end{align}
that also includes the Yang-Mills vertex for $s=1$ (again, Chan-Paton factors assumed).

The second possibility works \textit{only in three dimensions} for $s_1=s_2=s$. It has three derivatives for any $s$. We shall call it pseudo-minimal coupling to spin one field. It is given by the expression:
\begin{align}
\mathcal{V}_{s,s,1} = y_1 y_2 y_3 z_3^{s-1}\,,
\end{align}
and includes the $F^3$ self-interaction for Yang-Mills fields for $s=1$.
Note also, that this expression works in any dimensions only for $s=1,2$.
In higher dimensions, the number of derivatives rises with $s$ for this type of couplings, while in three dimensions spin one field addresses all the spins $s\geq 2$ in an equivalent manner, but clearly distinguishes them from spins $s=0,1$.

We notice, that the one-derivative minimal coupling to Yang-Mills is absent for fields with spin higher than one. This coupling is however present (see e.g. \cite{Gwak:2015jdo}), if we assume that the vector field is not Yang-Mills, but Chern-Simons, with free equations of motion given by flat (linearised) curvature condition.
We will discuss the Chern-Simons interactions together with parity odd cubic vertices in the forthcoming publication.

\subsection{Coupling to gravity}\label{Gravity}
This case is of utmost interest, since it accommodates the vertex of minimal coupling to gravity, the one that does not exist in higher dimensions (in four dimensions on-shell descriptions allow for minimal coupling \cite{Bengtsson:1986kh,Metsaev:1991nb,Ponomarev:2016lrm,Conde:2016izb}, while the Fronsdal description does not accommodate it).
This vertex allows for bypassing the Aragone-Deser problem of coupling HS fields to gravity, with the expense of deforming the gauge transformation of the gravitational field itself. This mechanism is similar to Supergravity situation, when the minimal coupling of Rarita-Schwinger field to gravity necessitates the fermionic transformation of the metric, thus leading to supersymmetric extension of the isometry algebra. Similarly, in three dimensions, coupling of any massless HS bosonic field (including massless spin-two fields) enforces enlarging the isometry algebra to (super)algebras that involve HS Killing tensors.

The minimal coupling to gravity is given by:
\begin{align}
\mathcal{V}_{s,s,2} = y_3\,z_3^{s-1}(s\, y_1 z_1+s\, y_2 z_2+y_3 z_3)\,,\label{MinimalCoupling}
\end{align}
which is gauge invariant due to identities \eqref{eq:threeDerivDDI}. Remarkably, this expression makes sense also for $s=0,1,2$. In the latter case, one has to use identities \eqref{eq:twoDerivDDI} to show that \eqref{MinimalCoupling} is equivalent to the conventional massless spin two self-interaction (Einstein-Hilbert) vertex, that is gauge invariant in any dimensions:
\begin{align}
\mathcal{V}_{2,2,2} =G^2=(y_1 z_1+y_2 z_2+y_3 z_3)^2\,.
\end{align}
We will study more general cases in the following.

\section{General case}\label{General}

It is instrumental to take a careful look at vertices with given number of derivatives, starting from the lower derivative cases. Since there are no Schouten identities with one derivatives, it is immediate to see, that there are no vertices with zero derivatives, except for scalar self-coupling $\phi^3$, which is not constrained by gauge invariance.

\subsection{One derivative vertices}

We are going to look for general solution for a cubic TT vertex with one derivative, assuming $s_1\geq s_2\geq s_3\geq 1$ ($s_3=0$ case was completely covered above). One-derivative interactions are only possible if the sum of the spins in the vertex is odd.
We distinguish following two cases\footnote{Note, that for odd sum of the spins, the triangle inequalities cannot be saturated, while only one (minimal) violation of triangle inequalities is possible, allowing one-derivative ansatz for the vertex.}:
\begin{itemize}
\item
\textbf{Triangle inequalities are not satisfied: $s_1=s_2+s_3+1$.}

The only possible vertex monomial in this case is:
\begin{align}
\mathcal{V}_{s_1,s_2,s_3} =y_1 z_2^{s_3} z_3^{s_2}\,,
\end{align}
which is gauge invariant only for $s_2=s_3=0$ --- cubic vertex of scalar electrodynamics covered in \eqref{s00}.

\item
\textbf{Triangle inequalities are satisfied: $s_i<s_{i-1}+s_{i+1}$.}

The general ansatz would be:
\begin{align}
\mathcal{V}_{s_1,s_2,s_3} =(\alpha y_1 z_1+\beta y_2 z_2+\gamma y_3 z_3)z_1^{n_1}z_2^{n_2}z_3^{n_3}\,,\\
n_{i-1}+n_{i+1}+1=s_i
\end{align}
taking variations and equating to most general expression in terms of identities \eqref{eq:twoDerivDDI}, we get a unique solution:
\begin{align}
\alpha=\beta=\gamma\,,\qquad n_1=n_2=n_3=0\,\,\, \rightarrow\,\,\, s_1=s_2=s_3=1\,,
\end{align}
which corresponds to the familiar Yang-Mills vertex:
\begin{align}
\mathcal{V}_{1,1,1} =\alpha \,G\,.
\end{align}

\end{itemize}

We conclude, that there is no one-derivative vertex, if at least one of the spins is bigger than one.

\subsection{Two-derivative vertices}

Now we will turn to the case of two-derivative vertices. This is applicable to cases when the sum of the spins in the vertex is even. We will assume now $s_1\geq s_2\geq s_3\geq 2$, since the cases $s_3=0,1$ were already covered above.
We consider separately cases, based on the relations between spins.
\begin{itemize}

\item
\textbf{ Strict triangle inequalities are satisfied: $s_{i+1}+s_{i-1}\geq s_i+2$}.

Using \eqref{eq:twoDerivDDI}, we can reduce the number of monomials in the most general ansatz for the two-derivative TT vertex\footnote{It is straightforward to show using the identities \eqref{eq:twoDerivDDI} that the basis of independent monomials, second order in $y_i z_i$ is given by three monomials $y_{i+1} z_{i+1} y_{i-1} z_{i-1}\sim y_i z_i G$.}
\begin{align}
\mathcal{V}_{s_1,s_2,s_3} =(\alpha_1 y_1 z_1+\alpha_2 y_2 z_2+\alpha_3 y_3 z_3)G z_1^{n_1} z_2^{n_2} z_3^{n_3}\,.
\end{align}
The gauge variations of this vertex will be three derivative expressions. It is not hard to see, that using the identities \eqref{eq:twoDerivDDI} and \eqref{eq:threeDerivDDI} one can bring any monomial of these variations into a form, containing all three $y$'s\footnote{It is straightforward to show using \eqref{eq:twoDerivDDI} and \eqref{eq:threeDerivDDI} that the only independent monomial of third order in $y_i z_i$'s is $y_1 z_1 y_2 z_2 y_3 z_3$.}
\begin{align}
\delta_1 \mathcal{V}=[\alpha_1(n_2-n_3)+(\alpha_2-\alpha_3)(n_2+n_3+1)]y_1 y_2 y_3 z_1^{n_1+1} z_2^{n_2} z_3^{n_3}\,,\\
\delta_2 \mathcal{V}=[\alpha_2(n_3-n_1)+(\alpha_3-\alpha_1)(n_3+n_1+1)]y_1 y_2 y_3 z_1^{n_1} z_2^{n_2+1} z_3^{n_3}\,,\\
\delta_3 \mathcal{V}=[\alpha_3(n_1-n_2)+(\alpha_1-\alpha_2)(n_1+n_2+1)]y_1 y_2 y_3 z_1^{n_1} z_2^{n_2} z_3^{n_3+1}\,,
\end{align}
The condition of gauge invariance with respect to all variations has a unique solution, up to overall constant:
\begin{align}
\alpha_i=n_{i-1}+n_{i+1}+1=s_i-1\,,
\end{align}
therefore the vertex can be written as:
\begin{align}
\mathcal{V}_{s_1,s_2,s_3} =[(s_1-1) y_1 z_1+(s_2-1) y_2 z_2+(s_3-1) y_3 z_3] G z_1^{n_1} z_2^{n_2} z_3^{n_3}\,,\label{2vertex}\\
n_i=\tfrac12(s_{i-1}+s_{i+1}-s_i)-1\geq 0\,,\qquad\qquad\qquad
\end{align}
which involves minimal coupling to spin two \eqref{MinimalCoupling}. This can be made manifest, rewriting the vertex \eqref{2vertex} in a different, but equivalent form:
\begin{align}
\mathcal{V}_{s_1,s_2,s_3} =y_3 z_1^{n_1} z_2^{n_2} z_3^{n_3+1}[(s_2+s_3-2) y_1 z_1+(s_3+s_1-2) y_2 z_2+(s_3-1) y_3 z_3]\,.
\end{align}
For $s_3=2,\, s_1=s_2=s$, this reproduces the minimal coupling to gravity, given in \eqref{MinimalCoupling}.

We understood now that the minimal coupling to gravity is a part of a bigger family of two-derivative vertices in three dimensions, that exist for every triple of integer spins greater than one, with even sum and satisfying strong triangle inequalities. This result is in contrast with Metsaev classification in higher dimensions. Nevertheless it is not surprising, since the Chern-Simons theories of HS gravity, rewritten in the metric-like form, make use of two-derivative vertices \cite{Campoleoni:2012hp,Fredenhagen:2014oua,Gwak:2015jdo}.

\item
\textbf{Triangle inequalities are saturated: $s_1=s_2+s_3$.}

In this case the general ansatz for cubic vertices will be:
\begin{align}
\mathcal{V}_{s_1,s_2,s_3} =y_1[\alpha_1 y_1 z_1+\alpha_2 y_2 z_2+\alpha_3 y_3 z_3] z_2^{s_3-1} z_3^{s_2-1}\,,
\end{align}
which is only gauge invariant for $s_1=2\,,\, s_2=s_3=1\,,$ and $\alpha_1=\alpha_2=\alpha_3$, and describes the minimal coupling of Maxwell field to gravity.

\item
\textbf{Triangle inequalities are violated: $s_1=s_2+s_3+2$.}

In this case, the general ansatz for the vertex is:
\begin{align}
\mathcal{V}_{s_1,s_2,s_3} =y_1^2 z_2^{s_3} z_3^{s_2}\,,
\end{align}
and is not gauge invariant, unless $s_2=s_3=0\,,\, s_1=2$, which then describes standard scalar coupling to gravity (\eqref{s00} for $s=2$).

\end{itemize}

It is elementary to check that these cases cover all possibilities, since there are no candidate vertex monomials with two derivatives for $s_1>s_2+s_3+2$.

\subsection{Three-derivative vertices}

Now we turn to three-derivative vertices, that are minimal possible number in the case when $s_1\geq s_2\geq s_3 > 1$ and $s_1+s_2+s_3$ is odd. We again distinguish cases:

\begin{itemize}

\item
\textbf{Triangle inequality is satisfied: $s_1< s_2+s_3$.}

In this case, any three-derivative vertex monomial contains third order polynomials in $y_i z_i$'s, and can be uniquely written in the form, containing all the $y_i$'s (we omit the arbitrary coupling constant in front):
\begin{align}
\mathcal{V}_{s_1,s_2,s_3} = y_1\, y_2\, y_3\, z_1^{n_1}\, z_2^{n_2}\, z_3^{n_3}\,,\\
n_{i-1}+n_{i+1}+1=s_i\,,\qquad\qquad
\end{align}
and is gauge invariant due to identities \eqref{eq:fourDerivDDI}. This vertex exists for any spins with odd sum $s_1+s_2+s_3$ and satisfying strong triangle inequalities:
\begin{align}
s_{i-1}+s_{i+1}>s_i\,,
\end{align}
Three derivative vertices $s-s-1$ and $(s+1) - s - 2$ are of this type.

\item
\textbf{Triangle inequality is minimally violated: $s_1=s_2+s_3+1$.}

In this case, the general ansatz for the vertex is given by:
\begin{align}
\mathcal{V}_{s_1,s_2,s_3} =y_1^2[\alpha_1 y_1 z_1+\alpha_2 y_2 z_2+\alpha_3 y_3 z_3] z_2^{s_3-1} z_3^{s_2-1}\,,
\end{align}
which is gauge invariant only for $s_2=s_3=1\,,\, s_1=3$, and coincides with \eqref{s-1-1}.

\item
\textbf{Triangle inequality is maximally violated: $s_1=s_2+s_3+3$.}

The vertex ansatz is given as:
\begin{align}
\mathcal{V}_{s_1,s_2,s_3} =y_1^3 z_2^{s_3} z_3^{s_2}\,,
\end{align}
and is not gauge invariant, unless $s_2=s_3=0$, coinciding with \eqref{s00} in the latter case.

\end{itemize}

\subsection{Vertices with higher number of derivatives}

Due to the four derivative Schouten identities \eqref{eq:fourDerivDDI} any non-trivial vertex term with $n\geq 4$ derivatives contains at least $n-1$ power of one of the $y_i$'s. Due to other identities that could be derived by combinations of all the Schouten identities \eqref{eq:twoDerivDDI}, \eqref{eq:threeDerivDDI} and \eqref{eq:fourDerivDDI}, we restrict even more the possible vertex monomials with high derivatives.

All candidate expressions for $s_1\geq s_2\geq s_3\geq 2$ with higher than three derivatives, satisfying weak triangle inequality $s_1\leq s_2+s_3$, can be shown to be equivalent up to Schouten identities to expressions, involving $y_1^2y_2y_3$, therefore are vanishing due to \eqref{eq:fourDerivDDI}. The only expressions that are not equivalent to zero up to Schouten identities, are those for $s_1>s_2+s_3$, and can be brought to the form $(n\geq 3)$:
\begin{align}
y_1^nz_2^mz_3^k\,,\quad y_1^nz_1z_2^mz_3^k\,,\quad y_1^ny_2z_3^k\,,\quad y_1^n (\alpha y_2z_2+\beta y_3 z_3)z_2^mz_3^k\,,
\end{align}
which all fail to be gauge invariant. One can carefully select all options, and see that all the gauge invariant cubic vertices with more than three derivatives are those with scalar and Maxwell fields.
We conclude, that there are no non-trivial interactions of fields with spins $s_1\geq s_2\geq s_3\geq 2$ with more than three derivatives. 
Since we already studied in detail scalar and Maxwell cases, 
this completes the classification of parity even vertices of massless fields in three dimensions.

\section{Discussion}\label{Discussion}

We have classified all parity even cubic interactions between massless bosonic fields in three dimensions. A remarkable difference of three dimensional vertices compared to higher dimensional ones is that for any three spins $(s_1,s_2,s_3)$ there is at most a unique vertex.
We will call massless fields of spin $0, 1$, scalar and Maxwell fields, matter fields, since they carry propagating degrees of freedom in three dimensions.
It is interesting to note, that the classification of cubic vertices in three dimensions departs from the classification in higher dimensions when there are more than one field of spin $s\geq 2$ in the vertex, or, in other words, there are less then two matter fields.
The vertices that coincide with the higher dimensional ones are all those containing at least two matter fields, spin two self-interaction (cubic order of linearised Einstein-Hilbert action, see e.g. \cite{Manvelyan:2010wp}) and spin three couplings $(3,s,s)$ with $s\leq 3$. One more curiosity of this classification is that spin three couples to all spins through $(s,s,3)$ couplings ``universally'' in three dimensions, similarly to spin two (the latter property is associated to equivalence principle) --- all of these vertices have three derivatives. The difference however is that it requires charged fields for that coupling.

The spin values, for which the vertex is absent in three dimensions  involve one matter field: $s_1\geq s_2\geq 2\,,\,\, s_3=0$ and $s_1> s_2\geq 2\,,\,\, s_3=1$. All the vertices that do not involve matter fields have two (three) derivatives for even (odd) sum of spins in the vertex. The only vertex with no derivatives is $\phi^3$ scalar self-interaction. The only vertices with one derivative are scalar coupling to Maxwell and Yang-Mills vertex. $(s,0,0)$ and $(s,1,1)$ vertices contain $s$ derivatives, $(s,1,0)$ vertex contains $s+1$ derivatives.

Each vertex in flat space can be uplifted to $(A)dS$ space (see e.g. \cite{Joung:2012fv,Francia:2016weg}). Therefore, via the AdS/CFT dictionary the classification of cubic vertices in flat space should conform to the structure of three point functions in 2d CFTs, which is indeed the case \cite{DiFrancesco:1997nk}. For full dictionary, one needs to derive also parity odd vertices, which is a work in progress.

Maxwell field in three dimensions is dual to a scalar, therefore it may be not so surprising that scalar and Maxwell fields admit similar current type couplings. One remarkable difference still exists: Maxwell field allows for $(s,s,1)$ coupling while there is no $(s,s,0)$ coupling for $s\geq 2$. Therefore, dual fields do not admit exact equivalence in their local interactions. It is worth to note, that if we have only one copy of each spin in the spectrum, $(s,s,1)$ interaction is trivial, therefore at the level of cubic vertices we will not see difference between scalar and Maxwell fields, but we believe that this difference is crucial in the full interacting theory, since dualisation is not a local operation.
We will discuss this dualisation aspect more in the forthcoming work, where we classify parity odd cubic vertices in three dimensions.

We notice also, that the scalar field interaction with higher spins \eqref{s00} requires higher derivatives.
It is a consistent vertex of current interaction type, contains $s$ derivatives and requires transformation of the scalar field of the following type:
\begin{align}
\delta^1_{\epsilon}\phi=\epsilon^{\mu_1\cdots\mu_{s-1}}\partial_{\mu_1}\dots\partial_{\mu_{s-1}}\phi\,,\label{ga}
\end{align}
where we suppress Chan-Paton factors, if any.

It can be shown that once one has a scalar field interacting with spin $s\geq 3$ by a cubic current interaction \eqref{s00}, we need to introduce infinite tower of HS fields, to close the algebra of gauge transformations \eqref{ga} of the scalar field. The argument is analogous to that of Berends,  Burgers and vanDam \cite{Berends:1984rq}.
It is argued in \cite{Joung:2014qya} that the theory with global symmetries containing finitely many HS killing tensors, cannot have a finite number of propagating degrees of freedom in $(A)dS)_3$ space, since the corresponding matrix algebras do not allow for unitary representations with GK dimension 2 (the number of independent continuous parameters, needed to describe a representation with finite species of particles, propagating in three dimensions). It remains to see if the special case of $sl_3\oplus sl_3$ algebra of the ``spin two plus spin three'' theory, introduced in \cite{Campoleoni:2010zq} can allow for scalar coupling. This case is singled out by having a unitary representation with GK dimension 2 and the fact that spin three cannot couple to a single real scalar by current coupling \eqref{s00}, therefore may avoid the aforementioned no-go arguments. If such a model exists, the corresponding matter has to carry trivial irrep of one of the $sl_3$'s and the minimal representation of the other $sl_3$. It is not clear if this representation can be realised field-theoretically. We leave this question to the future work.

\section*{Acknowledgements}

The author thanks Pan Kessel for stimulating discussions that led to the results presented here. Many useful discussions with Eduardo Conde Pena, Dario Francia, Euihun Joung and Gabriele Lo Monaco on the subject of this work are also gratefully acknowledged.
The author is also grateful to Eduardo Conde, Euihun Joung, Dario Francia, Eugene Skvortsov and Stefan Theisen for comments on the draft.
The author is also grateful to his wonderful hosts at Kaustinen during Christmas, the Teirikangas-Rytioja-Valo family, for the comfortable environment where this work has been completed.
This work is supported by Alexander von Humboldt Foundation.

\appendix


\end{document}